\documentclass[aps,prl,bibnotes,amsmath,amssymb,reprint,superscriptaddress]{revtex4-2}

\usepackage{bm}
\usepackage{graphicx}
\usepackage[colorlinks,
			linkcolor=blue,
            anchorcolor=blue,
            citecolor=blue,
            urlcolor=blue,
            breaklinks=true]{hyperref}

\newcommand{\eref}[1]{Eq.~(\ref{#1})}
\newcommand{\fref}[1]{Fig.~\ref{#1}}
\newcommand{\ie}{\emph{i.e.}}
\newcommand{\eg}{\emph{e.g.}}

\newcommand{\expv}[1]{\ensuremath{\langle #1 \rangle}}
\newcommand{\vg}{\ensuremath{v_\mathrm{g}}}
\newcommand{\leff}{\ell_{z,\mathrm{eff}}}
\newcommand{\zfc}{z_{\mathrm{fc}}}
\newcommand{\lone}{\ell_{z,\mathrm{eff}}^{(1)}}
\newcommand{\chitz}{\chi_{tz}}

\newcommand{\dd}{\mathrm{d}}

\newcommand{\MSD}{\mathrm{MSD}}

\begin{document}

\title{Temporal Branched Flow of Light}

\author{Weifeng Ding}
\thanks{These authors contributed equally to this work.}
\affiliation{Department of Physics, School of Science and Research Center for Industries of the Future, Westlake University, Hangzhou 310030, P. R. China.}
\affiliation{Institute of Natural Sciences, Westlake Institute for Advanced Study, Hangzhou 310024, P. R. China.}

\author{Chaokai Yang}
\thanks{These authors contributed equally to this work.}
\affiliation{Zhejiang Key Laboratory of Micro-nano Quantum Chips and Quantum Control, School of Physics, Zhejiang University, Hangzhou 310027, China}

\author{Zhaoying Wang}
\email{zhaoyingwang@zju.edu.cn}
\affiliation{Zhejiang Key Laboratory of Micro-nano Quantum Chips and Quantum Control, School of Physics, Zhejiang University, Hangzhou 310027, China}
\affiliation{State Key Laboratory of Ocean Sensing, ZJU-Hangzhou Global Scientific and Technological Innovation Center, Zhejiang University, Hangzhou 311215, China}

\author{Fanglin Bao}
\email{baofanglin@westlake.edu.cn}
\affiliation{Department of Physics, School of Science and Research Center for Industries of the Future, Westlake University, Hangzhou 310030, P. R. China.}
\affiliation{Institute of Natural Sciences, Westlake Institute for Advanced Study, Hangzhou 310024, P. R. China.}


\begin{abstract}
Branched flow of light arises when spatially correlated disorder folds families of rays into caustics. Here, we demonstrate a temporal analog in the retarded-time coordinate of an optical pulse in a waveguide, where branched flow becomes hyper-dispersive. Retardation during pulse propagation induces spatiotemporal correlations that break propagation reciprocity and imprint a trajectory memory. This nonreciprocity enables unidirectional temporal branched flow with no spatial counterpart. A memory model is developed combining ray-density and phase fluctuations with single-ray dynamics in a memory cell. The model reveals competition between fluctuation-induced focusing and destruction, which yields nontrivial diagrams of propagation regimes with distinct dispersion scalings. For a 532-nm pulse in a $\mathrm{LiNbO_3}$ fiber, we predict that refractive-index disorder induced by active modulations or quantum vacuum fluctuations can produce first caustics on meter-to-kilometer scales, offering routes to experimental observation and to sensing ultra-weak fluctuations.
\end{abstract}

\maketitle

The temporal degree of freedom offers a new perspective on exploring wave phenomena \cite{Zaletel2023} through its analogy with spatial dimensions. Temporal photonics with time-varying refractive indices has led to the discovery of temporal reflection and refraction \cite{Mendonca2000, Dong2024}, time lenses \cite{Kolner1989}, temporal cloaking \cite{Fridman2012}, periodic time crystals \cite{Lyubarov2022, Saha2023, Liu2023, Lustig2023, Asgari2024, Guo2026}, and temporal Anderson localization (TAL) in disordered media \cite{Carminati2021, Apffel2022}. Yet, between ordered modulation and uncorrelated disorder lies a nontrivial intermediate regime actively studied in spatial wave physics: branched flow \cite{Topinka2001, Barkhofen2013, Patsyk2020Nature, Garnier2025}, where correlated disorder produces tree-like propagation patterns with a hyper-diffusion scaling law \cite{Kaplan2002, Metzger2010, Metzger2014}. Here we ask whether light waves can branch in time, resulting in ``hyper-dispersion".

Conventional spatial branched flow (SBF) of light is governed by an effective Schr\"odinger equation \cite{Patsyk2020Nature}: propagation distance acts as the evolution variable, while caustics form across transverse spatial coordinates \cite{Patsyk2022}. The associated refractive-index landscape is typically modeled as a quasi-static random potential with equal longitudinal and transverse correlation lengths and no mutual cross-correlation between them \cite{Chang2024BF, *[{During the preparation of this manuscript, related work appears on anisotropic spatial correlations, see, \eg,~}]Chang2026}. A temporal analog would require time to serve as the effective transverse space, whereas propagation distance remains the effective time. This shift brings temporal envelope dynamics to the forefront and introduces a crucial distinction: during propagation, retarded time mixes propagation distance with laboratory time. The analog thus maps spatial disorder not to pure time disorder \cite{Kim2023}, but to spatiotemporal disorder with anisotropic correlation scales and nonzero cross-correlations, leading to novel features beyond existing SBF.
\begin{figure}[tb]
    \centering
    \scalebox{1}{\includegraphics{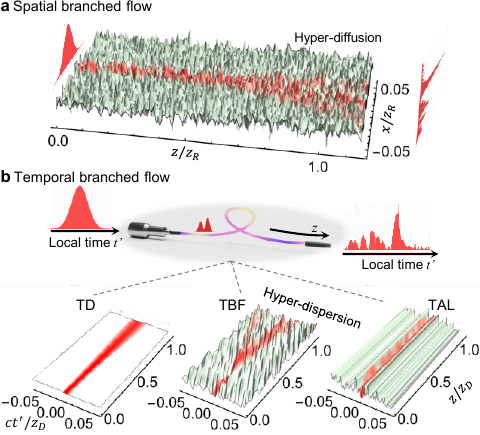}}
    \caption{From spatial to temporal branched flow: numeric illustrations. (a) Light beam branching in a 2D surface. Rayleigh distance: $z_\mathrm{R}=5.1\,\mathrm{mm}$. Potential fluctuation: $\sigma_n=0.001$. Correlation length: $\ell_c=20\,\mu\mathrm{m}$. (b) Optical pulse branching in a 1D waveguide, where the local retarded time becomes the configuration coordinate, and the propagation distance serves as the evolution variable. TD: temporal dispersion ($\sigma_n=0$). TBF: temporal branched flow ($a_{tz}=0.02$). TAL: temporal Anderson localization ($a_{tz}=0$). For TBF and TAL: $\sigma_n=0.005$, $c\ell_t=20\,\mu\mathrm{m}$, $\ell_z=200\,\mu\mathrm{m}$. Dispersion length: $z_\mathrm{D}=10.2\,\mathrm{mm}$.}
    \label{fig:schematics}
\end{figure}

In this Letter, we demonstrate temporal branched flow (TBF) of an optical pulse propagating in a waveguide (\fref{fig:schematics}) and develop a memory model for the underlying spatiotemporal ray dynamics. The model identifies a competition between ray-density fluctuation-induced focusing and ray-phase fluctuation-induced destruction, which drives crossovers from TBF to TAL or ordinary temporal dispersion (TD, \ie, group-velocity dispersion). Because spatial and temporal disorder are not equivalent, these crossovers form nontrivial propagation-regime diagrams. Retardation-induced spatiotemporal correlations further break propagation reciprocity, giving rise to unidirectional TBF.

\paragraph*{Temporal Schr\"odinger equation.}
Consider an optical pulse in a time-varying, one-dimensional waveguide with a real, disordered refractive index, $n(t,z)=n_0-\delta n(t,z)$, where $n_0 = \expv{n}$ is the mean index, $|\delta n|\ll n_0$, $z$ is the propagation distance, and $t$ is the laboratory time. The electromagnetic field of the pulse takes a paraxial-like form, $E(t,z)=A(t,z)e^{i(\omega_0t-k_0z)}+\mathrm{c.c.}$, with $A$ being the slowly varying envelope, $\omega_0$ the carrier frequency, $k_0=n_0\omega_0/c$, $c$ the vacuum speed of light, and c.c. denoting the complex conjugate. Neglecting back reflection, the electromagnetic wave equation reduces to an effective Schr\"odinger equation (see the Supple. Mat. for a full derivation),
\begin{equation}
i\partial_zA =-\frac{k''}{2}\partial_{t'}^2A
-\frac{k_0}{n_0}\delta n\!\left(t'+\frac{z}{\vg},z\right)A,
\label{eq:schrodinger}
\end{equation}
which is in the retarded time, $t'=t-z/\vg$, rather than in the laboratory time. Here, $\vg$ is the group velocity, $k''=\partial_\omega^2k|_{\omega_0}$ the group-velocity-dispersion coefficient, and $k(\omega)$ the propagation constant. Note that in the above equation, space and time have exchanged roles: $z$ is the evolution variable, while $t'$ is the configuration coordinate. The equation is thus termed the temporal Schr\"odinger equation, where dispersion supplies the kinetic term, and the refractive index fluctuation serves as the potential. 

In the laboratory coordinate, we model the disorder correlator with a Gaussian function,
\begin{equation}
    \begin{split}
        C_\mathrm{lab} &= \expv{\delta n(t,z)\delta n(t+\Delta t,z+\Delta z)}\\
        &= \sigma_n^2\exp\!\left[-\frac{\Delta t^2}{2\ell_t^2}
-\frac{\Delta z^2}{2\ell_z^2}
-C_{tz}\frac{\Delta t\,\Delta z}{\ell_t\ell_z}\right].
    \end{split}
\end{equation}
Here, $\sigma_n$ is the fluctuation strength, $\ell_t$ is the correlation time, and $\ell_z$ is the correlation length. The intrinsic cross-correlation satisfies $|C_{tz}|<1$, ensuring a negative Gaussian exponent. Changing to the retarded time, the correlator becomes
\begin{gather}\label{eq:M}
    C_\mathrm{ret}(\Delta t',\Delta z) = \sigma_n^2\exp\{-(\Delta t',\Delta z)\mathbf M(\Delta t',\Delta z)^\mathsf{T}/2\},\nonumber\\
    \mathbf M = 
    \begin{pmatrix}
    1/c^2\ell_t^{2} & a_{tz}/c\ell_t^{2}\\
    a_{tz}/c\ell_t^{2} & 1/\leff^{2}
    \end{pmatrix},\\
    \leff^{-2} = a_{tz}^2\ell_t^{-2}+(1-C_{tz}^2)\ell_z^{-2},\nonumber\\
    a_{tz}=1/\vg + C_{tz}\ell_t/\ell_z.\nonumber
\end{gather}      
Retardation during pulse propagation contributes to the overall spatiotemporal correlation through $1/\vg$. When propagation direction reverses, this retardation-induced contribution in $a_{tz}$ flips sign, whereas the intrinsic contribution captured by $C_{tz}$ is fixed, leading to propagation nonreciprocity. The overall spatiotemporal correlation tilts the potential landscape in the retarded space-time coordinates by an angle $\theta$, with $\tan 2\theta=2\mathbf{M}_{12}/(\mathbf{M}_{11}-\mathbf{M}_{22})$; see Fig.~S1 in the Supple. Mat. 

\begin{figure}[tb]
    \centering
    \scalebox{1}{\includegraphics{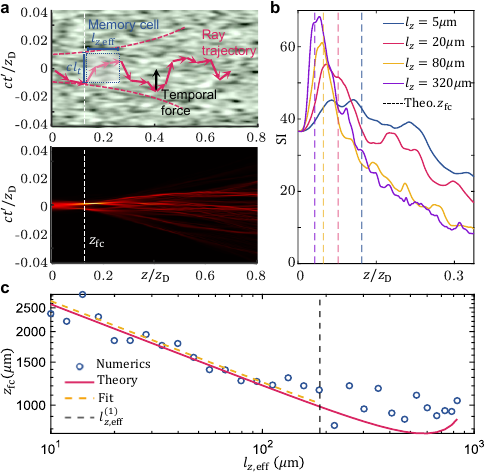}}
    \caption{(a) Dynamics of a temporal ray (red arrow) under random temporal force (black arrow) predicts the first-caustic distance (white dashed line). Top: potential landscape. Bottom: TBF. Red dashed curves: dispersion region (exaggerated for clarity). Blue dotted rectangle: memory cell. (b) Scintillation index curves under different correlation lengths, in comparison with theoretical first-caustic distances. Each curve is obtained by averaging over 20 runs. $c\ell_t=10\,\mu\mathrm{m}$. (c) Scaling of the first-caustic distance $z_\mathrm{fc}$ with the memory length $\leff$ shows a nonlinear region ($\leff > \lone=187\,\mu\mathrm{m}$) where memory strength decreases significantly. $\sigma_n=0.6\times10^{-3}$, $a_{tz}=0.01$, $z_\mathrm{D}=10.2\,\mathrm{mm}$.}
    \label{fig:bf}
\end{figure}

\paragraph*{The memory model.}
A temporal ray propagating from ($t'_0,0$) to $(t',z)$ traces a specific trajectory, $t'=T'(t'_0,z)$, in the retarded space-time coordinates (see \fref{fig:bf}a). The effective correlation length $\leff$ defines a memory length and, together with the correlation time, spans a memory cell of the spatiotemporal disorder. This motivates us to propose a memory model for the ray dynamics. We construct the optical field by interference,
\begin{equation}
    A(t',z) = \int A_{0}(\alpha) \cdot \delta[t'-T'(\alpha,z)] \cdot e^{i\phi_\alpha(t',z)}\,\dd\alpha,
\end{equation}
where $\delta$ is the Dirac delta, $A_0$ is the input amplitude and $\phi_\alpha$ is the accumulated phase for ray $\alpha$. The mean-squared displacement (MSD) of the optical field is of primary interest,
\begin{equation}
    \MSD(z) = \expv{\expv{(t')^2}_{I}-\expv{t'}^2_{I}}_{\delta n}.
\end{equation}
Here, $I(t',z)=AA^*$ is the intensity distribution, $\expv{x}_{I} = \int xI\,\dd t'$, and $\expv{\cdot}_{\delta n}$ denotes the ensemble average over all possible realizations of the random potential.
After manipulations, we have
\begin{equation}\label{eq:model}
    \begin{split}
        \MSD &= \int A_0^2\cdot \expv{ [T'_0-t'_0]^2\cdot \sum_{\alpha} \frac{1}{J_\alpha}e^{i(\phi_0-\phi_\alpha)} }_{\delta n} \dd t'_0\\
        &\sim C_\mathrm{M}\cdot \underbrace{\expv{(T'_0)^2}_{\delta n}}_{\text{single-ray}}\cdot \underbrace{\expv{\frac{1}{J}}_{\delta n}}_{\text{ray-density}} \cdot \underbrace{\expv{e^{i\Delta\phi}}_{\delta n}}_{\text{phase}},
    \end{split}
\end{equation}
where, in the first equality, we have assumed uniform $A_0$ for simplicity, $T'_\alpha$ abbreviates $T'(\alpha,z)$, $\alpha\in \{\alpha: T'_\alpha=T'_0\}$ labels all temporal rays reaching $(T'_0,z)$, and $J_\alpha=|\partial T'/\partial \alpha|$ is the Jacobian depicting the inverse of the ray density. In fact (shall be shown below), $\phi$ depends on $\delta n$, $T'$ depends on $\partial_t\delta n$, while $J$ depends on $\partial^2_t\delta n$. They are thus statistically independent to leading order, enabling separation in the ensemble average. Furthermore, for rays $\alpha$ far from $t'_0$, their contributions vanish due to dephasing. This leads to a phenomenological memory model shown in the second line, where the first average captures the single-ray random walk. The latter two averages capture ray-density and phase fluctuations and shall be evaluated in a memory cell. The compensation factor $C_\mathrm{M}$ accounts for the integral, summation, and other approximations.

Our main result is that, after subtracting the disorder-free TD contribution,
\begin{equation}
\begin{gathered}
    \MSD(z) = \frac{(k'')^2C_\mathrm{M}D_\omega}{3}\cdot \mathcal{B}(z) \cdot z^3, \\
    \mathcal{B}(z) = \leff\cdot \eta(z), \\
    \eta(z) = 
 \underbrace{\chitz}_{\text{tilt}}\cdot\;
 \underbrace{e^{-S_h}}_{\text{focusing}}\cdot\;
 \underbrace{e^{-S_s(z)}}_{\text{destruction}},\label{eq:memorystrength}
\end{gathered}
\end{equation}
where $D_\omega = \sqrt{2\pi}(k_0\sigma_n/n_0\ell_t)^2$ is the frequency-domain dispersion coefficient, $\mathcal{B}$ is the branching factor, and $\eta$ is the memory strength satisfying $0<\eta<1$. The memory strength in \eref{eq:memorystrength} has three contributions corresponding to the three mechanisms in \eref{eq:model}, and it is key to the rich TBF phenomena we explain in detail below.

\paragraph*{I. Single-ray random walk and the first caustic.}
For a given ray, the Hamiltonian in \eref{eq:schrodinger}, $H=k''\omega^2/2-(k_0/n_0)\delta n$, yields the canonical equations,
\begin{equation}\label{eq:rays}
    \frac{\dd t'}{\dd z}=k''\omega,\qquad
 \frac{\dd\omega}{\dd z}=\frac{k_0}{n_0}\partial_{t'}\delta n,
\end{equation}
where $\omega=\partial_{t'}\phi$ is the conjugate frequency (see the Supple. Mat. for derivations). This leads to a random temporal force (acceleration) exerted on the ray along the time axis, $F(t'_0,z)=\partial_z^2t'=k''k_0(\partial_{t'}\delta n)/n_0$. It follows that the longitudinal force correlation is
\begin{equation}
\begin{aligned}
 K_F(s)&=\langle F(t'_0,0)F(t'_0,s)\rangle_{\delta n}\\
 &=\left(\frac{k''k_0}{n_0}\right)^2\frac{\sigma_n^2}{\ell_t^2}
 \left(1-\frac{a_{tz}^2s^2}{\ell_t^2}\right)
 e^{-s^2/(2\ell_{z,\mathrm{eff}}^2)}.
\end{aligned}
\label{eq:force-kernel}
\end{equation}

\begin{figure}[tb]
    \centering
    \scalebox{1}{\includegraphics{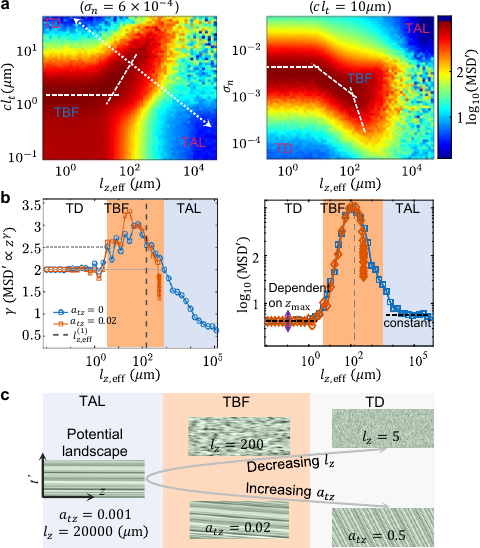}}
    \caption{(a) Crossover diagrams connecting TD, TBF, and TAL propagation regimes, averaged over 5 runs. (b) Cross section (right panel) along the white dotted arrow in (a) and the corresponding dispersion-scaling exponent (left panel). (c) Evolution of potential landscapes. The theoretical MSD peak $\lone$ is shown as white dashed lines in (a) and vertical gray dashed lines in (b). $n_0=2.28$, $w_0/c=50\,\mathrm{fs}$, $\lambda=633\,\mathrm{nm}$, $|k''| = 491.6\,\mathrm{ps}^2/\mathrm{km}$, $z_\mathrm{D}=10.2\,\mathrm{mm}$, and $z_\mathrm{max}=20\,\mathrm{mm}$.}
    \label{fig:gamma}
\end{figure}

According to \eref{eq:rays}, temporal index gradient changes frequency, and group-velocity dispersion converts that change into the ray's arrival-time deflection. The accumulated deflection assuming zero initial slope, $T'(0,z)=\int_0^z(z-u)F(u)\,\dd u$, yields a cubic scaling law,
\begin{equation}\label{eq:tbf}
    \begin{split}
        \expv{(T'_0)^2}_{\delta n} &= 2\int_0^z\!K_F(s)
        \left(\frac{z^3}{3}-\frac{z^2s}{2}+\frac{s^3}{6}\right)\dd s\\
        &\approx \frac{(k'')^2D_\omega}{3}\cdot \leff\cdot\chitz\cdot z^3,
    \end{split}
\end{equation}
which holds in the limit $z\gg \leff$, with
\begin{equation}\label{eq:chi}
    \chi_{tz}=1-\left(\frac{a_{tz}\ell_{z,\mathrm{eff}}}{\ell_t}\right)^2
 =\frac{(1-C_{tz}^2)\ell_{z,\mathrm{eff}}^2}{\ell_z^2}.
\end{equation}
Compared to disorder-free TD, where $\MSD_0=w_0^2(1+z^2/z_\mathrm{D}^2)$, $w_0$ is the input pulse duration and $z_\mathrm{D}$ is the dispersion length, we call the cubic law in \eref{eq:tbf} hyper-dispersion, analogous to the hyper-diffusion in SBF. Numeric simulations in \fref{fig:bf}a-bottom further demonstrate this hyper-dispersive propagation with tree-like patterns, \ie, the TBF; see the Supple. Mat. for simulation details. The first equality in \eref{eq:chi} implies $\chitz< 1$, and the second equality implies $\chitz>0$. $\chitz$ reduces the overall dispersion coefficient by tilting the potential landscape, as illustrated in \fref{fig:gamma}c.

The first caustic is expected when MSD reaches the temporal disorder correlation scale. By setting $\MSD(\zfc)=\ell_t^2$, we have the first-caustic distance,
\begin{equation}
 z_{\mathrm{fc}}=
 \left[
 \frac{3\ell_t^4}
 {\sqrt{2\pi}C_\mathrm{M}(k''k_0/n_0)^2\sigma_n^2\,
 \cdot\ell_{z,\mathrm{eff}}\cdot\eta}
 \right]^{1/3}.
 \label{eq:first-caustic}
\end{equation}
When memory strength is near unity, this implies $z_{\mathrm{fc}}\propto\sigma_n^{-2/3}$ and $z_{\mathrm{fc}}\propto(\ell_{z,\mathrm{eff}})^{-1/3}$, and reveals the non-equivalent roles of spatial and temporal correlation scales. \fref{fig:bf}b-c numerically verifies the scaling of $\zfc$ with respect to $\leff$. We locate $\zfc$ with the maximum of the scintillation index curve in \fref{fig:bf}b, $\mathrm{SI}(z)=\langle I^2\rangle_{t'}/\langle I\rangle^2_{t'}-1$, where the average is over the retarded time. We then show these maxima as circles in \fref{fig:bf}c, compared with \eref{eq:first-caustic}. For theoretical predictions, we set $C_\mathrm{M}=1$ and $\eta\to \chitz$. In the TBF regime (before the dashed vertical line), the $-1/3$ scaling is met. Fitting the data to the theoretical model (yellow dashed) yields $C_\mathrm{M}=0.91$ in this case. After the dashed vertical line, both numerics and theory show scaling-law deviations, implying that the memory strength decreases significantly.

\paragraph*{II. Ray density and fluctuation-induced focusing.}
To study memory strength and the crossover from TBF to other propagation regimes, we now evaluate the effects of ray-density and phase fluctuations. Differentiating \eref{eq:rays} with respect to input local time gives
\begin{equation}
\begin{gathered}
 \frac{\dd^2J}{\dd z^2}=\kappa(z)J,\qquad
 \kappa(z)=\frac{k''k_0}{n_0}\partial_{t'}^2\delta n,\\
 J(z)=1+\int_0^z(z-u)\kappa(u)J(u)\,\dd u.
\end{gathered}
\label{eq:jacobian}
\end{equation}
Here, we consider the initially unchirped ray family, with $J(0)=1$ and $\partial_z J(0)=0$. $\kappa$ represents the temporal curvature of the refractive index, with $\expv{\kappa(z)}=0$ and $\expv{\kappa(z)^2} \propto\expv{(\partial_{t'}^2\delta n)^2}\propto \sigma_n^2/\ell_t^4$. According to the above recurrence relation, we expand $J(z)$ within a memory cell in a Neumann series to the second order,
\begin{equation}
    \expv{J}_{\delta n}=1+\frac{\expv{\kappa(z)}\ell_{z,\mathrm{eff}}^2}{2}
 +\frac{\expv{\kappa(z)^2}\ell_{z,\mathrm{eff}}^4}{24},
\end{equation}
which yields
\begin{equation}
    S_h = -\ln\expv{1/J}_{\delta n} \sim C_h\frac{|k''|^2k_0^2\sigma_n^2}{n_0^2\ell_t^4}
 \ell_{z,\mathrm{eff}}^4,
\end{equation}
with a coefficient $C_h$. A memory cell has zero net temporal curvature, unlike a time lens ($\expv{\kappa}\neq 0$), but 2nd-order ray-density fluctuations effectively induce ray focusing, reducing the overall pulse spreading.

\paragraph*{III. Phase and fluctuation-induced destruction.}
The disorder-dependent phase accumulation is
\begin{equation}
    \phi_\alpha=\frac{k_0}{n_0}\int_0^z\delta n[t'_\alpha(u),u]\,\dd u.
\end{equation}
For $\Delta\phi_{\alpha\beta}=\phi_\alpha-\phi_\beta$, $\expv{\Delta\phi_{\alpha\beta}}=0$, the leading Gaussian cumulant gives
\begin{equation}
 e^{-S_s} = \left|\left\langle e^{i\Delta\phi_{\alpha\beta}}\right\rangle\right|
 =\exp\!\left[-\frac{1}{2}
 \left\langle(\Delta\phi_{\alpha\beta})^2\right\rangle_\mathrm{c}\right],
 \label{eq:phase-cumulant}
\end{equation}
where the subscript $\mathrm{c}$ denotes the connected variance. Unlike 1st-order destructive interference ($\expv{\Delta\phi_{\alpha\beta}}\neq0$), 2nd-order ray-phase fluctuations induce destructive pulse spreading. In the limit $z\gg \leff$, we have
\begin{equation}
    S_s \sim C_s (\frac{k_0}{n_0})^2\sigma_n^2 \leff\cdot z,
\end{equation}
with a coefficient $C_s$; see the Supple. Mat. for details.

\paragraph*{Propagation regimes.}
We now analyze propagation with the memory model. Varying the memory length, the branching factor peaks when $\partial\ln\mathcal{B}/\partial\leff=0$, yielding
\begin{equation}\label{eq:peak-general}
    1+\frac{\dd\ln\chi_{tz}}{\dd\ln\leff} = \frac{\dd S_h}{\dd\ln\leff} + \frac{\dd S_s}{\dd\ln\leff}.
\end{equation}
The optimal memory length $\lone$ that maximizes MSD is governed by the interplay among these three contributions to the memory strength. When potential tilt ($\chitz$) dominates, we have $\lone=\ell_t/\sqrt{3}a_{tz}$. When fluctuation-induced focusing ($S_h$) dominates, $\lone\propto\ell_t/\sqrt{\sigma_n}$. When fluctuation-induced destruction ($S_s$) dominates, $\lone\propto 1/z\sigma_n^2$. \fref{fig:gamma}a shows these analytical scalings as white dashed lines, consistent with the simulated, normalized MSD, where $\MSD'=\MSD/w_0^2$.
\begin{figure}[tb]
    \centering
    \scalebox{1}{\includegraphics{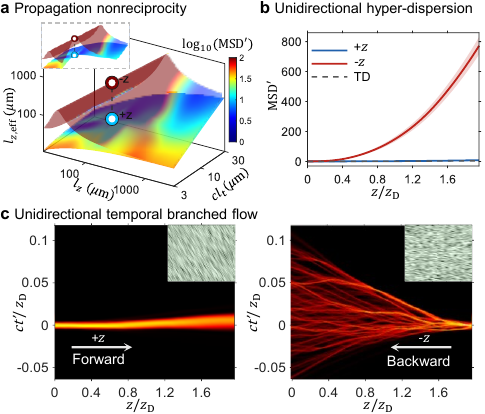}}
    \caption{For given spatiotemporal correlations ($\ell_t,\ell_z$), forward ($+z$) and backward ($-z$) propagation yield different memory-length sheets ($\leff$), leading to propagation nonreciprocity (a) and unidirectional TBF (c) with hyper-dispersion (b). For the illustrated case, $a_{tz}^{+}=0.188$, $a_{tz}^{-}=-0.008$, $c\ell_t=10\,\mu\mathrm{m}$, $\ell_z=100\,\mu\mathrm{m}$, $\sigma_n=0.6\times10^{-3}$, $z_\mathrm{D}=10.2\,\mathrm{mm}$; corresponding potential landscapes are shown in the insets of (c).}
    \label{fig:oneway}
\end{figure}

Denote the MSD scaling $\gamma=\dd\ln\MSD/\dd\ln z$. Around the MSD peak, we observe $\gamma=3$ for TBF (\fref{fig:gamma}b). In the limit $\leff\to \infty$, focusing dominates over destruction, the branching factor $\mathcal{B}\to 0$, and TBF crosses over to TAL ($\gamma=0$). When the observation position $z_\mathrm{max}\gg \leff$, the $z$-dependence of $S_s$ lowers $\gamma$ below 3. For small $\leff$ and large $\ell_t$, destruction dominates over focusing, and TBF crosses over to TD ($\gamma=2$). Empirically, we set $\gamma=2.5$ as the TBF-TD boundary, and $\gamma=2$ as the TBF-TAL boundary. Fitting the simulation data with the memory model, we obtain $C_s=1.03\times10^{-3}$ and $C_h=37.2$. During the crossovers, scanning $\leff$ by $a_{tz}$ or $\ell_z$ are equivalent but have different potential landscapes (\fref{fig:gamma}c).

\paragraph*{Experimental proposals.}
To observe TBF, we consider an optical pulse at $\lambda=532\,\mathrm{nm}$ propagating in a lithium-niobate fiber. The fiber's refractive index is modulated by quantum vacuum fluctuations through lithium niobate's nonlinearity in the Terahertz range. This modulation mechanism has been proven effective for sensing vacuum-field correlations \cite{Settembrini2022Vacuum}. Taking phase refractive index in Terahertz $n_{\mathrm{THz}}=2.23$, Pockels coefficient $r_{33}=31\,\mathrm{pm/V}$ \cite{Rabiei2004LNOI, Li2022VisibleLN}, and fluctuating field strength $E_{\mathrm{vac}}^{\mathrm{eq}}=\sqrt5\,\mathrm{V/m}$, we have $\sigma_n= n_{\mathrm{THz}}^3r_{33}E_{\mathrm{vac}}^{\mathrm{eq}}/2\approx 3.9\times10^{-10}$. For terahertz virtual photons, we take $\ell_t=1\,\mathrm{ps}$ and $\ell_z=300\,\mu\mathrm{m}$. With $C_{tz}=0$, $n_0=2.46$, and $|k''|=7.4\times10^3\,\mathrm{ps^2/km}$ \cite{Kaushalram2019Dispersion}, \eref{eq:first-caustic} gives $z_{\mathrm{fc}}\approx 400\,\mathrm{m}$. This proposal not only offers an example for experimental observation of TBF, but its first-caustic distance also provides a route to quantitatively sensing ultra-weak fluctuations. We note that a related pulse-breakup phenomenon has been observed in kilometers-long ultrahigh-PMD fibers \cite{Ning2003}, showing the feasibility of our proposed experiment.

Alternatively, active modulation can further enable characterization of the propagation-regime diagrams. The memory model is invariant under the following scale transformations, $\ell_t\to R\ell_t$, $|k''|\to R|k''|$, $\ell_{z,\mathrm{eff}}\to R\ell_{z,\mathrm{eff}}$, $z\to Rz$, and $\sigma_n\to\sigma_n/R$. With $R\sim10^3$ and $a_{tz}=0$, one can use experimentally achievable 10-GHz modulation to implement meter-scale TBF.

\paragraph*{Unidirectional TBF.}
Spatiotemporal disorder cross-correlation results in propagation nonreciprocity. For propagation along $+z$ and $-z$ directions,
\begin{equation}
 a_{tz}^{(\pm)}=\frac{1}{\vg}\pm C_{tz}\frac{\ell_t}{\ell_z}.
 \label{eq:direction}
\end{equation}
This leads to different $\ell_{z,\mathrm{eff}}^{(\pm)}$ and $\chi_{tz}^{(\pm)}$ for fixed $\ell_t$, $\ell_z$, and $C_{tz}$ (see \fref{fig:oneway}a), and hence allows different forward/backward propagation regimes. Note that even in the same propagation regime, distinct first-caustic distances satisfying $z_{\mathrm{fc}}^{(-)}<z_{\max}<z_{\mathrm{fc}}^{(+)}$ also allow one-way branching in practice. \fref{fig:oneway} demonstrates unidirectional TBF. For the selected configuration (dots) shown in \fref{fig:oneway}a, forward propagation has $\log_{10}\MSD'\approx0.9$, nearly in the TD regime, whereas backward propagation has $\log_{10}\MSD'\approx2.8$, branching in time with hyper-dispersive tree-like patterns.

\paragraph*{Discussion and conclusion.}
The real potential used in this work conserves the envelope norm $N=\int|A|^2\dd t'$, but the mean detuning obeys $\partial_z\bar\omega=(k_0/n_0)\langle\partial_{t'}\delta n\rangle_I$, and thus the pulse frequency is generally not conserved. For the tilted Gaussian correlator, the leading frequency drift is odd under $a_{tz}\to-a_{tz}$, whereas the first-caustic distance is even. This process is distinct from nonlinear pulse breakup, which usually requires gain, loss, self-phase modulation, or soliton fission \cite{Tai1986, Herrmann2002, Dudley2006}.

In summary, this Letter demonstrates temporal branched flow of light, which is the folding of the ray manifold in an optical pulse's retarded time. A memory model is developed to understand TBF and its crossovers to TD and TAL. The model reveals competition between fluctuation-induced focusing and destruction, which yields nontrivial propagation-regime diagrams. Compared with SBF, TBF features spatiotemporal cross-correlation, leading to unidirectional branched flow with no spatial counterpart. This work reveals a new class of spatiotemporal random-wave dynamics, with possible applications in sensing ultra-weak fluctuations.

\bibliographystyle{apsrev4-2}
\bibliography{temporal_branched_flow_references}

\end{document}